\documentclass[review]{elsarticle}

\usepackage{geometry}
\usepackage{graphicx}      
\usepackage{amsmath}
\usepackage{amssymb}
\usepackage{arydshln}

\usepackage{hyperref}

\journal{Journal}

\begin{document}

\begin{frontmatter}

\title{Assessing the different aspects of consuming fashion and the role of self-confidence on the buying behaviour of fashion consumers in the clothing market as a mediator}



%
\author[mymainaddress]{Milad Zam\corref{mycorrespondingauthor}}
\cortext[mycorrespondingauthor]{Corresponding author}
\ead{m.zam@ut.ac.ir}

\author[mysecondaryaddress]{Mohammadhosein Tavakoli}
\ead{mohammadhosein.tavakoli@alumni.esade.edu}

\author[mythirdaryaddress]{Hasan Ramezanian}
\ead{h_ramezanian@yahoo.com}

\author[myfourtharyaddress]{Amin Rezasoltani}
\ead{Aminrezasoltani123@gmail.com}

\address[mymainaddress]{Faculty of Management, University of Tehran, Tehran, Iran}
\address[mysecondaryaddress]{ESADE Business School, Ramon Llull University, Barcelona, Sant Cugat, Spain}
\address[mythirdaryaddress]{Department of Mechanical Engineering, Amirkabir University of Technology, Tehran, Iran}
\address[myfourtharyaddress]{Ava and Nima Social Robotics Co. (Dr. Robot), Tehran, Iran}

\begin{abstract}
As a mediator variable, self-confidence is one of the most effective elements of the decision-making process of consumer behaviour. 
This research has studied the effects of different aspects of consuming fashion on the self-confidence and behaviour of consumers in Tehran’s clothing market. This study has considered the acceptance of new products, interest in mode and fashion, utilitarianism, and personal taste in its analysis. 
This research aims to understand the fashion buying behaviour amongst Iranian consumers in consideration of their attitude towards self-confidence and aspects of fashion consumption. 
The statistic sample is 400 consumers from Tehran’s clothing market who have been chosen based on the random availability procedure. 
The primary tool in this research was a questionary used to testify the assumptions and a model fit created by using structural equations and factor analysis.
This research showed that the interest in mode and fashion, personal taste, utilitarianism, and new products positively impact self-confidence. In addition, the positive impact of self-confidence on fashion buying behaviour was confirmed. 

\end{abstract}

\begin{keyword}
Interest in modem personal taste, utilitarianism, new fashion products, self-confidence, fashion buying behaviour 
\end{keyword}

\end{frontmatter}


\section{Introduction}

These days, fashion is a vital industry with billions of dollars in revenue and has created jobs for millions of people worldwide~\cite{le2014fashion}. The pioneers and leaders of this industry hold various exhibitions and festivals every year and use celebrities to endorse their products to attract more people to this industry. 
Iran has also observed a change in clothing style in recent decades. As such, Iranian clothing producers have unintentionally fallen in competition with more internationally acceptable fashion and clothing styles.

Various reasons are helping to change people’s tastes and attitudes towards fashion. For example, Turkish and European brands, which are the leaders in this industry, or the red-carpet fashion awards in the US, show many celebrities worldwide and can be treated as fashion leaders by their fans. 
The attention and expansion of the fashion industry have been appreciated in Iran in recent years. This has occurred by promoting the industry by producers of top Iranian brands such as Hakoupian, Doris, Celebon etc, teaching and training people by leaders in the fashion and textile industry, holding seasonal exhibitions and festivals, and encouraging top producers to move toward the international market. 
However, they have not been able to compete as much and as planned with countries that are the leaders in this industry. The marketing style in clothing and attention to domestic consumer needs were keys to develop and expand in this domain.

\section{Problem formulation}

"\textit{In England, to do future planning in marketing, it is essential to understand fashion consumers}" needs, requirements, and demands. As each consumer is unique, it is crucial to use various segmentation, targeting, and positioning techniques to place them in rational groups. This method enables us to ensure that each unique consumer is getting offered the right fashion product~\cite{le2014fashion, tajdari2022feature}.
The fashion structure consists of people and organisations creating symbolic meanings and participating in conveying these meanings to cultural products. Fashion code or language is helping us in decoding these meanings. Fashion is a social diffusion process where consumers accept new styles and products~\cite{bin2021marketing, yang2021posture}. 

Fashion is a complex topic that works on different levels. On the one hand, fashion is a social phenomenon that simultaneously affects various groups of people. Furthermore, fashion is entirely personal, affecting an individual’s behaviour. Their motivation about fashion governs the consumer’s decision-making~\cite{bin2021marketing, esfandiari2021fuzzy}. Western fashion styles have influenced the Iranian market. However, The Iranian~\cite{zerafati2022multi} domestic market has lost the competition because the existing marketing knowledge in this area is not  sufficient. Also we witness no significant attention in analysing Iranian consumer behaviours, their attitude toward clothing,fashion, and mode. Fashion has a definition amongst buyers and companies, which conveys their figures and identities. Therefore, the product itself is beyond its character because the values related to the clothing will be added to the clothing~\cite{le2014fashion}. The expression fashion is a representation of clothing. However, fashion and mode apply to a broad spectrum of products such as mobile phones, cars, interior design, toys, musical instruments, etc. \cite{le2014fashion}
Fashion and fashionism take people’s feelings regardless of the time, location, culture and ethnicity. Research on style and context of fashion and analysis of consumer behaviour has always been an area of interest among designers and producers of clothing and other consumer products. The concept of fashion is related to people being unique in their attitudes, behaviours and lifestyle~\cite{rasay2021developing}. The appearance of some brands such as Dior, Armani Gucci, etc has emphasised their consumer’s tastes and tried to create particular styles to convey these unique feelings to their consumers and somehow promote their social level by using their products.

Clothing was used to protect human beings for the first time. Quickly, fashion took over as a leader from other aspects and characteristics of clothing, and people started to interpret and connect with it by using the latest fashion clothing. For example, in the 14th century, fashion became a representation of people’s social situation in western countries. From the mid-19th century, a significant expansion of factories in various industries occurred and that caused the expansion of distribution channels, the creation of larger clothing centres, artistic clothing shops, and helped some clothing brands to become internationally and commercially acceptable brands such as Dior, Gucci, Yves Saint Laurent and Chanel~\cite{le2014fashion}.  
It is essential in the fashion industry to understand consumer behaviour patterns~\cite{tajdari2019fuzzy, tajdari2020intelligent, tajdari2021simultaneouschanger}. Although there are 9266 official clothing retailers and 707 official clothing wholesalers in Tehran’s fashion industry, they cannot compete with other brands because they are unaware of the fashion buying behaviour of Iranian consumers. This research not only is trying to analyse fashion buying behaviour amongst Iranian consumers but also analysing self-confidence and its effect on consumer behaviour due to fashion buying behaviour, fashion information search, and new fashion products.

\section{The importance and necessity of this research}

The fashion industry is one of the newfound industries which is progressing rapidly. European countries mainly focus on analysing fashion processes and new fashion products that have directly caused development in this industry. It is needed to analyse the attitude of consumer behaviour because fashion products are complex and unique and can represent a type of feeling and unique identity of individuals~\cite{workman2017we}. Such analysis can be used to improve people's performance in this industry. In other words, when people are welcoming fashion, they would prefer to do something against the norm~\cite{lin2012cognitive}. 

Clothing based on the latest model and fashion style is more important in youth than in elderlies. It can be said that people who are for the new style usually feel younger~\cite{lin2012cognitive}. This research aims to examine the effect of fashion buying behaviour, fashion information search and new fashion products on self-confidence and the effect of self-confidence on the fashion industry as a leader. This subject has already been studied in other countries but has not been significantly explored in Iran. This research analyses the importance of fashion and mode based on its position within the community. The structural equations have been used to understand whether self-confidence is rolling as an intermediary between fashion buying behaviour, new fashion products and fashion leadership. The primary purpose of this research is to analyse the effect of being interested in fashion, accepting new fashion products, personal taste, utilitarianism over self-confidence and the role of self-confidence as a mediator on consumer behaviour.

\section{The research background}

Marketing people classify their customers based on some characteristics and variables to provide services and products to them. These classifications or groups are affected by some reference groups that can affect consumer behaviour. The theory concentrated on consumer behaviour expresses that consumers refer to people from different groups to seek help to shape attitudes, beliefs, and related behaviours regarding the selection and buying and commercial brands. The marketing people are willing to assess the consumer classifications to the reference groups and the membership of these consumers based on the group's level of compromise and influence on each individual~\cite{jackson2008mastering}. There is a significant distinction between the new generation and the old one. The new generation is more rebellious, whereas the older generation is more conservative. Therefore, grouping based on the type of generation provides more opportunities to market people and companies in the fashion industry because when the current generation becomes old, they try to use the products that represent their youth time~\cite{jackson2008mastering}. The usage of fashion products, notably clothing, demonstrates people’s attitudes toward themselves and their surroundings. The fact that has distinguished generations is the difference between the behaviour on how to use fashion and mode. This fact is essential for marketing in the fashion industry because the current generation buys different things from the older generation. In addition, there are always some consumers within their own generation and group behaving differently in clothing. In the history of fashion marketing research, most researchers believe that the leaders of fashion and mode are young consumers~\cite{morgan2009investigation}.

\subsection{the consumer’s behaviour and utilitarianism}

Consumers are looking for two goals in their consuming behaviours:
One is the pleasure from using it, which gives an excellent experience to the consumer (e.g. shopping for entertainment), and the other is the rational behaviour based on the needs. Researchers believe that there are various motives for shopping. From utilitarianism, the consumers are motivated to earn the product at a reasonable price. Consumers are motivated to get entertained by entertainment and pleasure irrespective of having any plans for any specific shopping~\cite{workman2010fashion}. In other words, it is more about spending time and enjoying the moments with people they like.

Consumer behaviour includes personal choice in buying and using some products to meet their needs. The need causes a reaction, but the reaction guides an individual’s behaviour. In reality, motivation is affected by physical, cognitive, emotional and social factors, which causes the behaviour to move in that direction. Therefore, motivation is one of the most critical topics in studying consumer behaviour. In addition, motivation is the consumer’s reaction to some particular products that meet their motivations. The motivations could be based on utilitarianism (practical and operational) or pleasure and entertainment (experimental and intuitive). From the utilitarian point of view, consumers are motivated to earn products at a reasonable price~\cite{workman2010fashion}. The usage of fashion products can bring an excellent experience to its consumers. Fashion is indeed a phenomenon that includes any products, from the latest technological products to clothing. Using a smart mobile phone, home appliances, music and celebrities’ clothes are some examples of fashion products. The usage of fashion products can affect the buying behaviour of consumers and their buying decision-making process.

\subsection{Fashion buying behaviour}
Fashion is a complex phenomenon that has various practical levels. On the one hand, fashion is a social phenomenon that simultaneously affects various groups of people. On the other hand, fashion is entirely personal, affecting an individual’s behaviour. Their motivation for fashion governs consumers’ decision-making~\cite{bin2021marketing}. The majority of consumer behaviour is based on their wishes, not their needs. In fact, the ideal is what the consumer is asking for regardless of whether they need it or not. Workman and Stoddart figured out that the behaviour of fashion fans is more likely to be based on their needs; the fashion leaders change their buying behaviour based on their wishes~\cite{workman2012gender}. Fashion consumers have different emotions and engagements with different types of products. Clothing is recognised as an attractive product and an emotional motivation. Therefore, clothing has an emotional element because it engages with an individual’s identity who has emotionally been induced~\cite{workman2012gender}.

\subsection{Fashion new product}

The product can be considered an answer to consumers' needs or a solution to their problems. Fashion is a complete tool to demonstrate being a social group member, and clothing is a typical tool to join specific reference groups or peer groups. The reason is that commercial brands and products are shining when they are used. The product of fashion is the most sector in fashion marketing~\cite{jackson2008mastering}. The fashion structure consists of people and organisations creating symbolic meanings and participating in conveying these meanings to cultural products. Fashion code or language is helping us in decoding these meanings. Fashion is a social diffusion process where consumers accept new styles and products~\cite{bin2021marketing}.

\subsubsection{Self-confidence, an interest in fashion and personal taste in fashion buying behaviour}

Howard believes that self-confidence is assurance that the buyer knows about their judgements toward a desirable or undesirable brand. Self-confidence demonstrates the consumer’s belief in knowing that their knowledge and abilities are sufficient about fashion clothing~\cite{o2004fashion}. They need to be unique and have the most effect on consumers' engagement with fashion clothing subjectively. This tendency can be appeared on buying fashionable clothing. One way to demonstrate uniqueness is to wear unique clothes and demonstrate the individual's personality. If people do not feel good about being unique, they will automatically try and buy clothing according to the latest fashion. This type of consumer needs more novelty and complexity. Therefore, they show more mental complexities with their choice of fashion clothing. Consumers with more mental complexity believe that they are getting admired when they wear a new style of cloth, and by doing so, they convey new information to others. They like to impress others through interpersonal conversations and their unique tastes in fashion~\cite{klepp2005reading}.

The clothing is very much under the influence of fashion. Fashion can be understood as a commercial system dictating the changes in appearance and clothing. Therefore, fashion is beyond the individual’s control and affects the type of clothing and the attitude and taste of the individual~\cite{klepp2005reading}.  Then people who think they are young are willing to try new products of different brands and are likely to do more research about fashion products. These people are confident and are likely to become leaders in changing attitudes toward fashion~\cite{nam2007fashion}. The interest in fashion is structurally essential to define the type of people who are more interested in the physical appearance of clothing~\cite{reisenwitz2007comparison}. In addition, the taste brings a level of enthusiasm toward fashion. The individual’s taste is part of self-concept~\cite{workman2017we}.

\section{The Assumptions and conceptual model of the research}

\begin{figure}[tb]
	\centering
	\includegraphics[width=\linewidth]{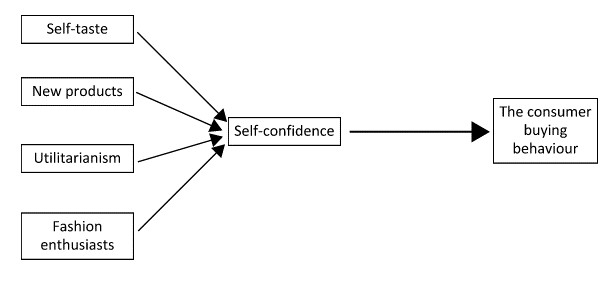}
	\caption{The conceptual model of self-confidence and consumer buying behaviour.}
	\label{fig:fig1}
\end{figure}

Self-confidence and product knowledge are not the same in consumers. Based on the situation, self-confidence can demonstrate certainty or uncertainty about whether the decision is to go with the self-confidence or product knowledge~\cite{workman2012gender}. In larger social structures, people have integrity in their behaviours and self-concepts, which assist them in valuing their self-concepts. This allows them to show a picture of themselves by using their behaviour and reflecting that in society. The primary motivation can be having a positive frame from an individual or maintaining an identity that talks about people wearing a different clothing style~\cite{mathur2005antecedents}. Previous research has also shown that self-perception about age can be related to behavioural variables such as the tendency to use new brands,\cite{stephens1991cognitive} searching data\cite{gwinner2001testing}, and fashion knowledge~\cite{wilkes1992structural, sudbury2009understanding}. Aston also believes that there is a key concept in all fashion theories, and that is that the leaders in fashion are those who accept changes in fashion quicker than the others and are fearless in using new styles. Therefore, they become an inspirational reference to the rest. The fashion followers support this industry by following it and considering the fashion industry leaders as their role models. \cite{bailey2010relationships}
Based on the previous research, we can assume the following:
\begin{enumerate}
\item An interest in fashion has a meaningful effect on self-confidence 
\item The new fashion products have a meaningful effect on self-confidence
\item The self-taste has a meaningful effect on self-confidence
\item Utilitarianism has a meaningful effect on self-confidence 
\item Self-confidence has a meaningful effect on the consumers’ behaviour
\end{enumerate}
All of the above show that people have different attitudes about themselves when they are using fashion products, showing their interest in fashion, tending to accept new fashion products or showing some utilitarianism behaviour. 
The conceptual model in Figure~\ref{fig:fig1} can be created based on the scholar’s findings of relevant factors to consumer behaviour, self-confidence and their effects on the fashion buying behaviour, and the above assumptions.

\section{Psychological Research}
This research is a correlation type of research when looking at its nature, purpose, and how the information has been collected. This research has used the correlation matrix analysis, which includes confirmatory factor analysis due to the existence of assumption, and has used structural equation modelling due to the examination of variables. The tool that has been used in this research is a structured questionary that was distributed amongst consumers in Tehran’s grand bazaar. The multiple-choice structure was used in two different structures: never, rarely, sometimes, most of the time and consistently, and the second one was strongly disagree, disagree, neither agree nor disagree, agree and strongly agree. 
In this research, the effect of self-confidence was considered from different aspects of using fashion. The statistical group of fashion consumers in the clothing industry in Tehran comprised 3,870,546 females and 3,876,087 males and included both married and single. In this research, the statistical group was divided into four age categories: 20-29, 30-39, 40-49 and 50-59 and each was further sub-divided based on gender. 
The random availability procedure was used in this research because the purpose of this research was to measure the variables of self-confidence, the leadership index of fashion in sampling and extending it to the society as a whole. Our statistical community is unlimited, so we used the Cochran formula for it. In this formula, $s^2$, as the largest variance, Z and d was calculated as 0.991, 1.96, and 0.1, respectively, where the sampling size was 384 people. 

In order to experiment, 30 questionnaires were randomly distributed in clothing retailers, and 24 of those were completed and collected. In order to measure the variance of the experimental questionary, we first calculated the Cronbach’s alpha by using SPSS 16 software and then calculated the average and variance of each variable. Finally, after computing the sampling size using the Cochran formula, 400 questionnaires were distributed amongst clothing consumers in Tehran. To do a reliability assessment of the questionary, Cronbach’s alpha was calculated, and those that had Cronbach’s alpha of less than 0.5 were removed. The result of Cronbach’s alpha of the variables of this research is shown separately in Table 1. 
Based on Table 1, the reliability of Cronbach’s alpha values of the questionary (above 0.6) was confirmed. Also, the questions were inherited consistent, which demonstrates the accuracy of answers by the participants. This research used the techniques of descriptive statistics and inferential statistics. In addition, correlation analysis was used to examine the assumption, and structural equation modelling based on the variances or partial least squares was used by SPSS 16 and Smart-Pls 3 software to assess the fashion consuming variable's effects on fashion buying behaviour with self-confidence as a mediator.

\subsection{Assessing findings, reliability and justification of the research structure}

The structural path modelling (variance approach) and the structural equation modelling (co-variance approach) are two main ways of analysing and assessing complex data structures~\cite{anwar2021influence, tajdari2021imageprediction, tajdari2020intelligentcontrol, ghaffari2018new, khodayari2015new, tarvirdizadeh2017assessment}. The main characteristic of these two methods is assessing and analysing different dependant and independent variables concurrently~\cite{tajdari2020semirobot, tajdari2021discrete, tajdari2021implementation, tajdari2022online}. Many factors were observed to assess the type of the research, such as the data of the 400 questionnaires, multipliers, determining the variance of the dependant variable and the indirect effect of all variables on each other. 
In this examination, the acceptable factor load loading of the variable for each variable is 0.7, and the p-value is 0.1. As shown in Table~\ref{tab:table1}, the factor loading of each variable is above 0.7, which proves the reliability of the data. 

Also, to assess the reliability of the data~\cite{tajdari2021adaptive, tajdari2022flowsaturation, tajdari2020feedback, tajdari2019integrated}, the composite reliability of each structure was used. The acceptable criterion for the reliability of the structure is a value above 0.7. As per Table~\ref{tab:table1}, the composite reliability of fashion buying is 0.829, utilitarianism is 0.829, new fashion product is 0.893, personal taste is 0.890, self-confidence is 0.894, and fashion enthusiast is 0.855. Therefore, it can be said that the composite reliability of the structure is acceptable. 
In order to assess the convergent validity of structures, the average of the structures of this research was computed by PLS. The acceptable criterion is the values above 0.5, and as per Table~\ref{tab:table1}, the average variance (AVE) of each structure is above 0.5. This variable is computed as 0.623 for utilitarianism, 0.546 for a new fashion product, 0.620 for personal taste, 0.597 for fashion enthusiasts, 0.739 for self-confidence, and 0.645 for buying behaviour

\begin{table}[tb]
	\renewcommand{\arraystretch}{1.3}
	\caption{the average variance (AVE)}
	\label{tab:table1}
	\centering
	\includegraphics[width=\linewidth]{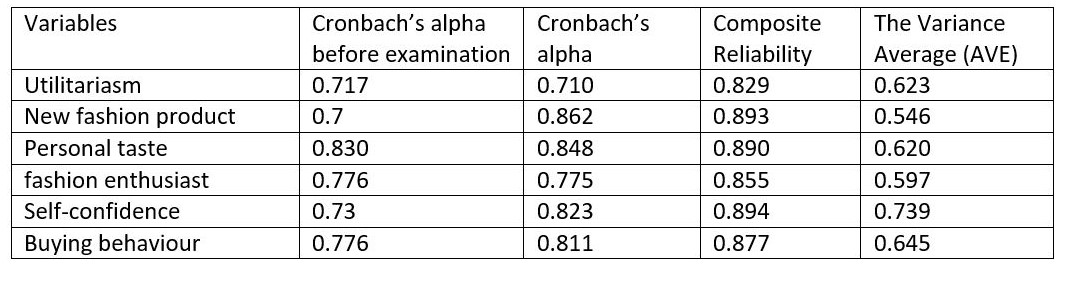}
\end{table}

In order to assess the correlation between the hidden variables \cite{golgouneh2016design, tajdari2017switching, tajdari2017design, tajdari2017robust}, the intended data were taken from AVE and Correlation Latent Variable tables. As the square root of the AVE is more than the correlation of one structure to another, the sampling model's diagnosis at the level of structure is justifiable. The values in Table~\ref{tab:table2} are the square root of the variance average (AVE)~\cite{minnoye2022personalized, tajdari2022optimal, tajdari2022flow, tajdari2022dynamic}. 

\begin{table}[tb]
	\renewcommand{\arraystretch}{1.3}
	\caption{the correlation between the hidden variables ( Fornel and Locker Analysis}
	\label{tab:table2}
	\centering
	\includegraphics[width=\linewidth]{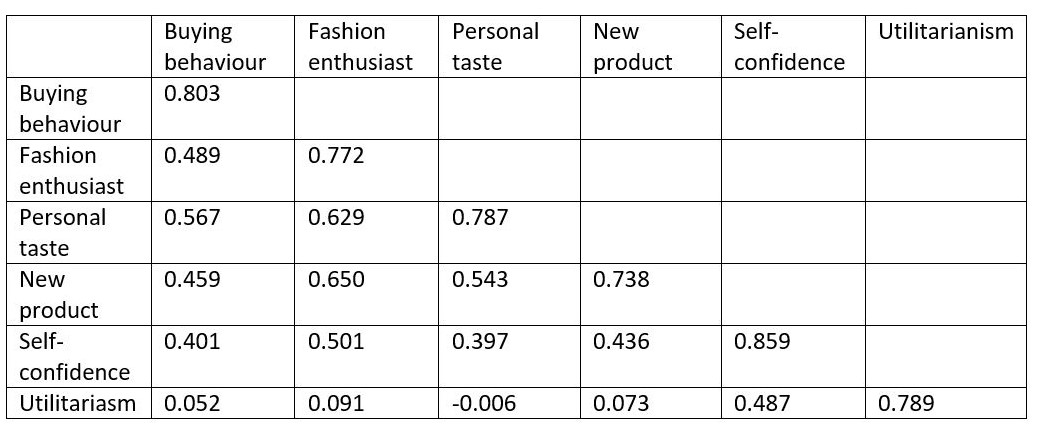}
\end{table}

\section{Assumptions examination}

\begin{figure}[tb]
	\centering
	\includegraphics[width=\linewidth]{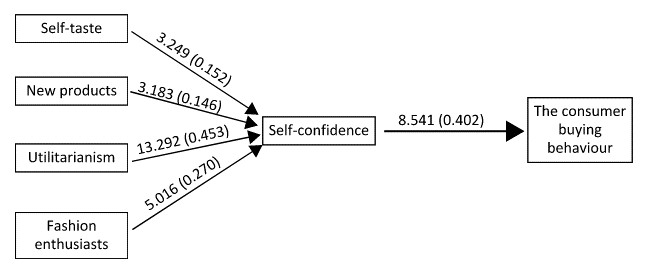}
	\caption{P-Value and Path Analysis}
	\label{fig:fig2}
\end{figure}

After the fit for purpose examination of the research model, the research assumptions were examined. The T-values and the statistics of each path analysis are shown in Figure\ref{fig:fig2} in each bracket. As long as the calculated value is within the assurance range, the assumption can be accepted or rejected. The P-value of this research is 95\%, and the T-Value is 1.96. Table~\ref{tab:table3} shows the final results of this research modelling

In order to assess the path analysis, the t-value was computed from the PLS software. One of the indexes that confirms the relations between the structural modellings is the beta multiplier. The z values are the incremental values, and the direction is the beta multiplier. The assumption will be accepted if the calculated value is within an assurance range. The P-values of 90\%, 95\% and 99\% were compared with the minimum t-value of 1.64, 1.96 and 2.58. Table~\ref{tab:table3} shows the final results of the modelling of this research.

\begin{table}[tb]
	\renewcommand{\arraystretch}{1.3}
	\caption{the final results of the modelling}
	\label{tab:table3}
	\centering
	\includegraphics[width=\linewidth]{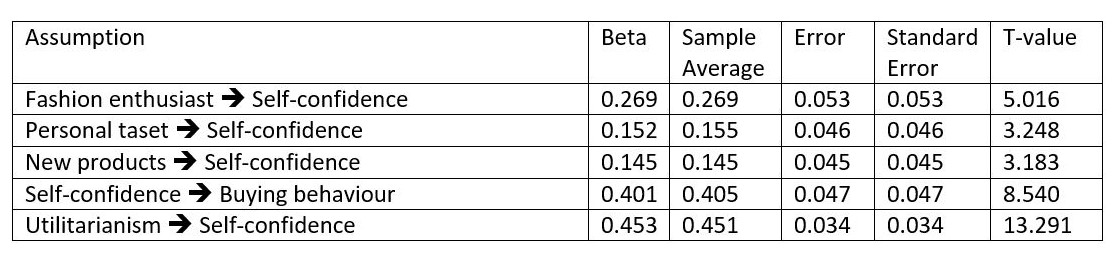}
\end{table}

The results of examining the assumptions and t-value are shown in Table~\ref{tab:table4}. These results confirm the accuracy of the assumptions. The P-value in the first assumption is 5.016, which shows a meaningful effect of an interest in fashion on self-confidence. In assessing the effects of personal taste, new products and utilitarianism on self-confidence, the t-value is 3.249, 3.183, 13,292, respectively. These values are more than 1.96, which shows that each variable significantly affects self-confidence. The t-value to assess the effects of self-confidence on consumers' buying behaviour is 8.541, which is greater than the t-values where the P-value is 95\%. Therefore, this assumption is also acceptable.

\begin{table}[tb]
	\renewcommand{\arraystretch}{1.3}
	\caption{The results of examining the assumptions and t-value where P-value is 95\%}
	\label{tab:table4}
	\centering
	\includegraphics[width=\linewidth]{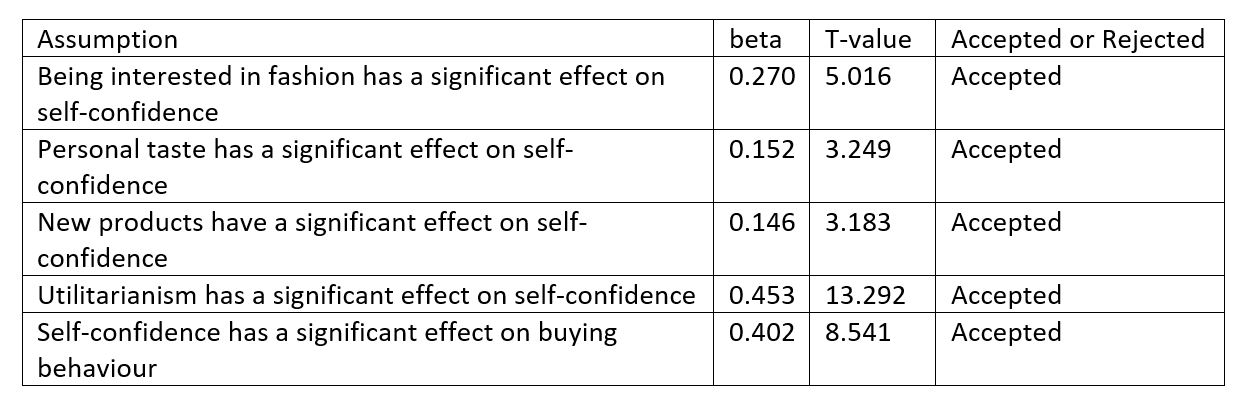}
\end{table}

\section{Conclusion}

One of the purposes of this research is to assess the effects of self-confidence as a mediator on a consumer's behaviour and their choice and usage of fashion in clothing. This paper has studied fashion from the self-confidence point of view due to the widespread usage of advertisement and marketing of the leaders in the fashion industry and the change in the buying behaviour of consumers in Iran. The research model was reviewed based on the attitude of people in Iran toward fashion and any previous research on this topic. We added a few more variables to this research to show a broader dimension of fashion that requires more assessment. 

As the calculated t-value is 8.541 and is larger than 1.96, it can be concluded that self-confidence is confirmed to have a significant effect on buying behaviour. These results have also shown that fashion buying can increase the self-confidence of fashion consumers, and because this value is positive, the effect of self-confidence on buying behaviour is considered positive. As explained in the background of this research, people in larger social communities are trying to maintain a unity between themselves and their behaviours to add value to their fantasies. The primary motivation can be having a positive picture or a preserved identity that leads to wearing different clothing styles. 
In addition, based on the outcome and data of this research from the descriptive statistical section, women feel younger than men, and therefore, young girls and women can be targeted for fashionism in Iran. Also, single people feel younger than married people. Therefore, their tendency to fashion products is more and a step to produce new products and create a new style in clothing can bring an opportunity to Iranian manufacturers within the Iranian clothing market~\cite{hadian2020practical}. 

People living in the North part of Tehran are willing to accept more than people living in the East part of Tehran, and selling new products may not be a good idea within the East part of Tehran. Therefore, it is better to target the North part of Tehran for fashion products. 
There were some constraints in preparing this research, such as lack of cooperation of some of the participants, inaccessibility to all retailers due to time, and lack of a more precise assessment of consumers' taste. The followings are suggested for the future researchers to take into their consideration due to the importance of marketing in the fashion industry as well as the importance of this industry for domestic manufacturers:
\begin{itemize}
\item The assessment of self-confidence in consuming fashion products can provide a different perspective on this aspect of consumer behaviour. Without a doubt, many factors can assist in providing self-confidence when buying products or searching for information that requires more attention to the consumer's behaviour.
\item Attention to topics such as branding has become more popular amongst fashion manufacturers and marketing people as opposed to the attention to the behaviour of consumers. This has caused fashion industry leaders to get away from a different aspect of consumer behaviour, and it is needed to carry out more research on the behaviour of fashion consumers.
\end{itemize} 











\bibliography{mybibfile}

\end{document}